\documentclass[runningheads]{llncs}
\usepackage[T1]{fontenc}
\usepackage{newtx}

\usepackage{graphicx}

\usepackage{booktabs}
\usepackage{multirow}
\usepackage{multicol}
\usepackage{booktabs}
\usepackage{bigstrut}
\usepackage{hyperref}
\usepackage{xcolor}
\usepackage{float}
\usepackage{credits}
\usepackage{orcidlink}
\usepackage{colortbl}
\begin{document}
\title{SENS3: Multisensory Database of Finger-Surface Interactions and Corresponding Sensations}
\titlerunning{SENS3 Database}

\author{Jagan K. Balasubramanian\orcidlink{0000-0002-8990-9629} \and
Bence L. Kodak\orcidlink{0009-0003-8557-7465} \and
Yasemin Vardar\orcidlink{0000-0003-2156-1504}}
\authorrunning{Balasubramanian, Kodak, and Vardar}

\institute{Delft University of Technology,\\
Mekelweg 2, 2628 CD Delft, The Netherlands\\
}
\maketitle              
\begin{abstract}
The growing demand for natural interactions with technology underscores the importance of achieving realistic touch sensations in digital environments. Realizing this goal highly depends on comprehensive databases of finger-surface interactions, which need further development. Here, we present \href{https://www.sens3.net}{SENS3}---\textit{www.sens3.net}--- an extensive open-access repository of multisensory data acquired from fifty surfaces when two participants explored them with their fingertips through static contact, pressing, tapping, and sliding. SENS3 encompasses high-fidelity visual, audio, and haptic information recorded during these interactions, including videos, sounds, contact forces, torques, positions, accelerations, skin temperature, heat flux, and surface photographs. Additionally, it incorporates thirteen participants' psychophysical sensation ratings (rough--smooth, flat--bumpy, sticky--slippery, hot--cold, regular--irregular, fine--coarse, hard--soft, and wet--dry) while exploring these surfaces freely. Designed with an open-ended framework, SENS3 has the potential to be expanded with additional textures and participants. We anticipate that SENS3 will be valuable for advancing multisensory texture rendering, user experience development, and touch sensing in robotics.

\end{abstract}
 \section{Introduction}

Recent trends in interactive system development emphasize a naturalistic approach to replicating human engagement with their physical surroundings in digital environments, promising enhanced user experience, accessibility, and improved digital communication~\cite{giri2021application}. While everyday tasks in the physical world involve engaging with objects through multiple senses~\cite{lederman2004multisensory}, transferring this rich sensory information to the digital domain remains a challenge despite advancements in hardware and algorithms for capturing and recreating real-world sensory experiences~\cite{xia2018new} and understanding multisensory integration~\cite{cornelio2021multisensory}. 

Besides device and algorithm design, achieving naturalistic human-technology interactions relies on one more essential element: data. Thanks to the advancements in camera and microphone technologies, one can effortlessly capture and share their surrounding's audio and visual information. Unfortunately, the same practicality does not apply to tactile data. When humans touch an object, they feel a rich array of tactile cues revealing its distinct surface properties, such as friction, roughness, thermal, and compliance~\cite{Okamoto2013PsychophysicalDO}, depending on the applied exploratory procedures, e.g., sliding, static contact, and pressing~\cite{lederman1987hand}. High-fidelity data collection for each property and exploratory procedure requires specialized expertise and recording technology. This situation makes it challenging for regular users to record the tactile feel of every encountered surface instantly. A fundamental issue is that tactile cues, unlike light and sound waves, require physical contact. The resulting skin deformations depend on finger and surface properties and applied normal force and speed~\cite{richardson2022learning}; these can vary substantially even for the same user and surface, hugely influencing the recorded tactile data~\cite{tanaka2014contact}. Due to these reasons, only a few available databases~\cite{culbertson2014one,strese2017content,strese2019haptic,richardson2022learning} include tactile recordings from interactions with surfaces. Moreover, existing databases often concentrate on single or dual sensory cues, e.g., vision and tactile, audition and tactile, or tactile only, insufficient for lifelike digitization of surfaces or neglect bare-finger interactions, the most natural way of interacting with our surroundings~\cite{lederman1999perceiving}. Finally, 
these databases mainly do not include psychophysical sensations humans perceive upon interaction such as rough--smooth, flat--bumpy, sticky--slippery, hot--cold, regular--irregular, fine--coarse, hard--soft, and wet--dry; providing such information can tremendously help understand human texture perception or generate algorithms for machine perception.

To address the above issues with the existing datasets, we propose a novel database, \href{https://www.sens3.net}{SENS3\footnote{\href{https://www.sens3.net}{www.sens3.net}}}, encompassing all necessary multisensory cues for naturalistic texture digitization. Our open-access database includes visual, auditory, and tactile data recorded while two participants explored 50 surfaces with their fingertips. Data collection measurements were conducted with a custom-designed apparatus, and they consisted of four exploratory procedures: static contact for thermal aspects, pressing and tapping to record compliance and hardness properties, and sliding to capture roughness and friction. Additionally, \href{https://www.sens3.net}{SENS3} includes psychophysical sensations rated by thirteen participants while freely interacting with the selected surfaces. We also introduce a user-friendly website that describes data collection procedures and collected data for a broad audience. We envision that our database will significantly impact various fields, such as human-machine interaction and robotics, providing a comprehensive and rich resource for researchers and developers to enhance the realism and authenticity of texture rendering and perception and, ultimately, to improve the user experience.

\begin{table}[!h]
\caption{Comparison of attributes of available haptic texture databases with \href{https://www.sens3.net}{SENS3}. Star, *, indicates a not freely accessible database.}
\centering
\resizebox{12.3cm}{!}{
\begin{tabular}{cccccccc}
\hline
Dataset                                                   & Audio/ Visual                                                    & Tactile                                                                                                          & Exploration                                                  & Interaction                                                                        & Sensations                                                      & \begin{tabular}[c]{@{}c@{}}Texture \\ count\end{tabular} & \begin{tabular}[c]{@{}c@{}} Participant \\ count\end{tabular} \\ \hline
 \begin{tabular}[c]{@{}c@{}}HaTT \\ ~\cite{culbertson2014one}\end{tabular}                                                     & Image                                                            & \begin{tabular}[c]{@{}c@{}}3D contact forces\\ \& accelerations,\\ finger speed\end{tabular}                     & Sliding                                                      & Tool-tip                                                                           & No                                                              & 100                                                           & 1            \\ \hline
\begin{tabular}[c]{@{}c@{}}CBSMR \\ ~\cite{strese2017content}\end{tabular}                                                   & \begin{tabular}[c]{@{}c@{}}Audio, image \\ \& video\end{tabular} & \begin{tabular}[c]{@{}c@{}}3D contact forces\\ \& accelerations,\\ reflectance,\\ conductivity\end{tabular}      & \begin{tabular}[c]{@{}c@{}}Tapping\\ \& sliding\end{tabular} & Tool-tip                                                                           & \begin{tabular}[c]{@{}c@{}}Pairwise\\ similarities\end{tabular} & 108                                                           & 1            \\ \hline
                                                          &                                                                  & 3D accelerations                                                                                                 & Sliding                                                      & \multirow{2}{*}{\begin{tabular}[c]{@{}c@{}}Bare-finger\\ \& tool-tip\end{tabular}} &                                                                 &                                                               &              \\ \cline{3-4}
                                                          &                                                                  & Reflectance                                                                                                      & \begin{tabular}[c]{@{}c@{}}Contour \\ following\end{tabular} &                                                                                    &                                                                 &                                                               &              \\ \cline{3-5}
  \begin{tabular}[c]{@{}c@{}}LMT \\ ~\cite{strese2019haptic}\end{tabular}                                                     & \begin{tabular}[c]{@{}c@{}}Audio \&\\ image\end{tabular}         & \begin{tabular}[c]{@{}c@{}}Thermal\\ conductivity\end{tabular}                                                   & Static touch                                                 & \begin{tabular}[c]{@{}c@{}}Standalone\\ sensors\end{tabular}                       & No                                                              & 184                                                           & 1            \\ \cline{3-5}
                                                          &                                                                  & 2D contact forces                                                                                                & \begin{tabular}[c]{@{}c@{}}Lateral\\ motion\end{tabular}     & \begin{tabular}[c]{@{}c@{}}Sensors \\ mounted\\ under\\ finger pad\end{tabular}    &                                                                 &                                                               &              \\ \cline{3-4}
                                                          &                                                                  & Pressure                                                                                                         & Pressing                                                     &                                                                                    &                                                                 &                                                               &              \\ \cline{3-5}
                                                          &                                                                  & Mass \& volume                                                                                                   & Holding                                                      & \begin{tabular}[c]{@{}c@{}}Standalone\\ sensors\end{tabular}                       &                                                                 &                                                               &              \\ \hline
 \begin{tabular}[c]{@{}c@{}}Haptex*\\ ~\cite{jiao2019haptex}\end{tabular}                                                   & No                                                               & \begin{tabular}[c]{@{}c@{}}3D contact forces\\ \& torques\end{tabular}                                           & Sliding                                                      & Bare-finger                                                                        & No                                                              & 120                                                           & 1            \\ \hline
\begin{tabular}[c]{@{}c@{}}Learn2feel\\ ~\cite{richardson2022learning}\end{tabular}                                              & Image                                                            & \begin{tabular}[c]{@{}c@{}}3D contact forces\\ \& accelerations,\\ finger speed\end{tabular}                     & \begin{tabular}[c]{@{}c@{}}Tapping\\ \& sliding\end{tabular} & Bare-finger                                                                        & \begin{tabular}[c]{@{}c@{}}Pairwise\\ similarities\end{tabular} & 10                                                            & 10           \\ \hline
\begin{tabular}[c]{@{}c@{}}Concurrent* \\ ~\cite{devillard2023con}\end{tabular}                                              & \begin{tabular}[c]{@{}c@{}}Image \\ \& video\end{tabular}        & \begin{tabular}[c]{@{}c@{}}3D contact forces\\ torques,\\ \& accelerations,\\ finger speed\end{tabular}          & Sliding                                                      & Bare-finger                                                                        & No                                                              & 10                                                            & 1            \\ \hline
\begin{tabular}[c]{@{}c@{}}Haptic \\ library~\cite{hassan2017towards}\end{tabular} & Image                                                            & No                                                                                                               &  \begin{tabular}[c]{@{}c@{}}Free\\ exploration\end{tabular}                                                           & Bare-finger                                                                        & \begin{tabular}[c]{@{}c@{}}Cluster\\ sorting\end{tabular}       & 84                                                            & 10           \\ \hline
                                                          &                                                                  & \begin{tabular}[c]{@{}c@{}}Skin temperature,\\ heat flux, 3D contact\\ forces \& torques\end{tabular}            & \begin{tabular}[c]{@{}c@{}}Static\\ contact\end{tabular}     &                                                                                    &                                                                 &                                                               &              \\ \cline{3-4}
\href{https://www.sens3.net}{SENS3}                                                     & \begin{tabular}[c]{@{}c@{}}Audio, image \\ \& video\end{tabular}         & \begin{tabular}[c]{@{}c@{}}3D contact forces\\ \& torques,\\ indentation depth\end{tabular}                      & Pressing                                                     & Bare-finger                                                                        & \begin{tabular}[c]{@{}c@{}}Adjective\\ ratings\end{tabular}     & 50                                                            & 2/ 13        \\ \cline{3-4}
\textbf{}                                                 &                                                                  & \begin{tabular}[c]{@{}c@{}}3D contact forces\\ \& accelerations\end{tabular}                                     & Tapping                                                      &                                                                                    &                                                                 &                                                               &              \\ \cline{3-4}
                                                          &                                                                  & \begin{tabular}[c]{@{}c@{}}3D contact forces,\\ torques \& accelerations\\ finger position \& speed\end{tabular} & Sliding                                                      &                                                                                    &                                                                 &                                                               &              \\ \hline
\end{tabular}
}
\label{tab:Datasets}
\end{table}

\section{Existing Haptic Texture Databases}\label{sec:related}

Several databases have been made to date for digitizing tactile textures; see Table~\ref{tab:Datasets} for a list of these prior work. 
The first comprehensive one, the Penn Haptic Texture Toolkit (HaTT), featured unconstrained tool-surface interactions with 100 surfaces, their models for haptic rendering, and high-resolution images of each surface~\cite{culbertson2014one}. The tactile data included contact forces, accelerations, and scan speed and was recorded with a custom-designed tool. The authors later demonstrated that the recorded contact accelerations could be used to re-create the vibratory feels of surfaces via a voice-coil actuator~\cite{culbertson2013generating}. 

HaTT mainly focused on roughness and friction modalities, overlooking capturing multisensory aspects of texture perception~\cite{lederman2004multisensory}. To address this gap, Strese \textit{et al.} developed two texture databases~\cite{strese2017content,strese2019haptic}. They recorded data using multiple custom recording devices while exploring more than 100 surfaces via sliding, tapping, contour following, static contact, pressing, and holding. Their recording devices allowed mainly tool- or sensor-surface interactions. The databases contained contact accelerations, forces, thermal conductivity, reflectance, mass, volume, and interaction sounds and videos; check Table~\ref{tab:Datasets} for detailed distribution of collected data in each database. 

Although tool/sensor-surface interactions were shown to effectively recreate the vibratory feel of surfaces and machine perception applications, the recorded data still does not represent the information obtained during bare-finger explorations~\cite{strese2019haptic}. To combat this, recent studies~\cite{jiao2019haptex,devillard2023con,richardson2022learning} introduced databases for finger-surface interactions. These databases include contact forces, accelerations, videos, and sounds recorded during sliding or tapping interactions; see Table~\ref{tab:Datasets} for more details of each dataset. The studies also showcased different uses of their data, such as bare-finger texture rendering~\cite{jiao2019haptex}, texture classification~\cite{devillard2023con}, or understanding human perception of surfaces~\cite{richardson2022learning}. Although these databases provide the much-needed information on finger-surface interactions, they only consider a subset of exploratory procedures and omit other important tactile properties, such as compliance and thermal. 

Finally, only some of the databases~\cite{vardar2019fingertip,richardson2022learning,hassan2017towards} include perception data along with the physical one, often focusing on pairwise similarities. While these similarities provide insights into how interaction data shapes human perception, articulating precise statements on how changes in contact data correlate with sensations remains challenging and necessitates adjective ratings for each sensation~\cite{richardson2022learning}.

\section{SENS3 Database}\label{sec:sens3}

We aimed to overcome the limitations of previous databases summarized in Section~\ref{sec:related} with \href{https://www.sens3.net}{SENS3}. We recorded the physical interaction data while participants explored the surfaces with their \emph{bare} fingers. As human fingertips show variety in geometry and mechanical properties~\cite{manfredi2014natural,serhat2022contact}, we collected this data from \emph{two} people. Moreover, as previous studies~\cite{lederman2004multisensory} showed evidence that all tactile, visual, and auditory cues contribute to surface perception, we recorded \emph{multisensory} information that spans these three senses. Additionally, we collected sensation ratings from \emph{13} people to complement the physical recordings to understand human texture and machine perception and develop algorithms for effective texture rendering.

For \textbf{tactile data}, we aimed to capture perceptually relevant multi-modal material properties: \emph{warmth, compliance, friction}, and \emph{roughness}~\cite{Okamoto2013PsychophysicalDO}. Previous works indicated that optimal exploratory procedures differ for the information to be gathered~\cite{lederman1987hand}. These procedures include lateral motion for roughness, pressure for compliance/hardness, and static contact for thermal properties. Furthermore, later studies showed that tapping is also suitable for hardness discrimination~\cite{richardson2022learning,culbertson2016importance}. Therefore, we collected information with four distinct exploratory procedures: static contact, pressing, tapping, and sliding (lateral motion). 

Because human fingertips are soft and show nonlinear behavior with applied pressure and moving speed, the interaction data also vary as a function of these parameters~\cite{tanaka2014contact}. Therefore, we collected contact forces and torques for all interactions. Due to the same reason, the data for the sliding experiments covered a spectrum of speeds (ranging from 0 to 200~mm/s) and applied forces (ranging from 0 to 1~N). Nonetheless, the rest of the interactions covered only one or two selected pressure ranges. In addition, we collected finger vibrations for active dynamic interactions (tapping and sliding), indentation depth for pressing, heat flux, and skin/surface temperature for static contact. See Table~\ref{tab:Datasets} for a detailed list of collected data in prior databases and \href{https://www.sens3.net}{SENS3}.

We took top-view high-resolution surface images for \textbf{visual data}. We also recorded dynamic finger-surface interactions with two cameras, one from the top and one from the side. Utilizing images and videos is essential, offering valuable perspectives from static and dynamic viewpoints. 
For \textbf{auditory data}, we recorded sounds of active dynamic finger-surface interactions (tapping and sliding). 

For the \textbf{perceptual data}, we recorded human participants' haptic psychophysical sensations, i.e., adjective ratings, resulting from their free multisensory (visual, audio, and haptic) exploration of the surfaces.

\begin{figure}[!b]
   \centering
    \includegraphics[width=\linewidth]{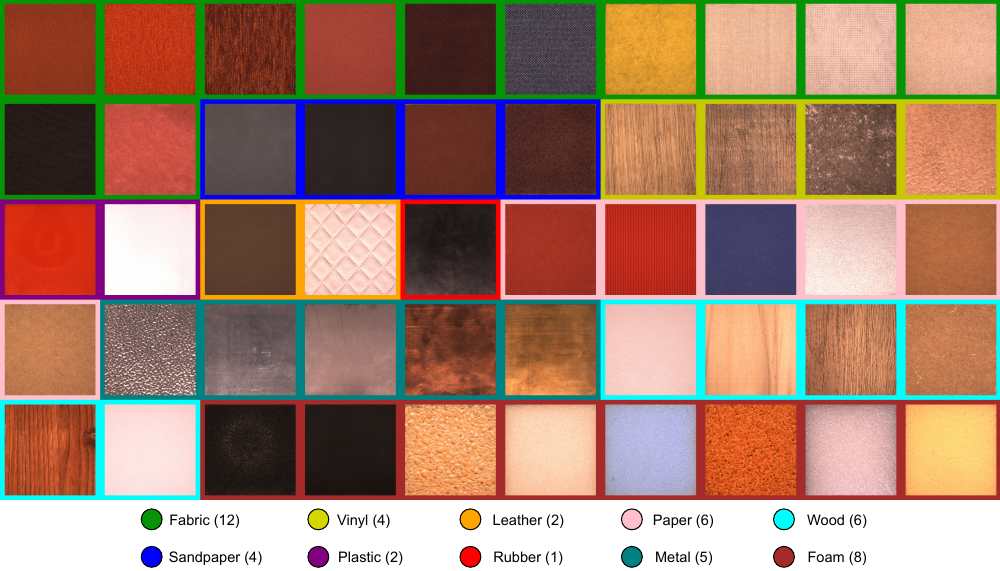}
    \caption{Recorded surfaces in \href{https://www.sens3.net}SENS3 database. The material categories are color-coded; each category's surface count is indicated in the brackets.}
    \label{fig:textures}
\end{figure}

\subsection{Selected surfaces \& data collection apparatus}
We selected 50 distinct homogeneous surfaces across \emph{ten} material categories: wood, metal, fabric, paper, rubber, plastic, sandpaper, leather, foam, and vinyl (see Fig.\,\ref{fig:textures}). These material categories encompass most of the encountered surfaces in daily life, and at least one texture is present per category. Each surface was cut to dimensions of 100$\times$100 mm and stuck on acrylic plates (3~mm thickness for each) using double-sided tape (Tesa PRO double-sided tape, 50mm$\times$25~m). 

\subsection{Data collection apparatus}

\begin{figure}[!h]
   \centering
   \includegraphics[width=\linewidth]{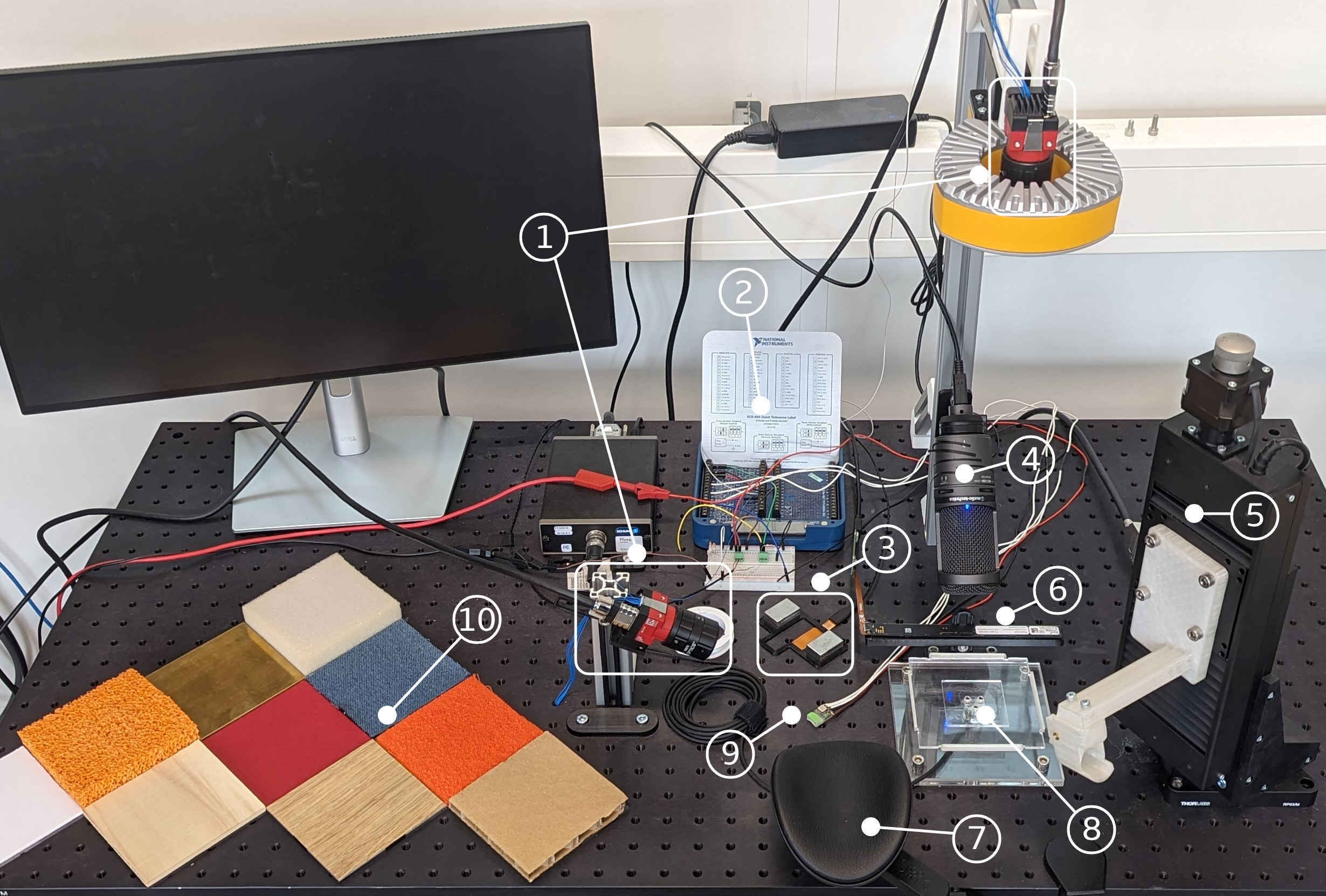}
   \caption{Data recording apparatus: 1. Cameras, 2. data acquisition board, 3. heat flux sensor, 4. microphone, 5. linear stage, 6. position sensor, 7. armrest, 8. force sensor, 9. accelerometer, 10. selection of surfaces.}
   \label{fig:Veni_HW}
\end{figure}

We built a custom data collection apparatus (see Fig.\,\ref{fig:Veni_HW} and Supplementary Video), which was placed on an aluminum optical breadboard table (MB60120/M, Thorlabs) and mounted onto a robust supporting frame (PFM52502, Thorlabs). During the experiments, a monitor (UltraSharp 24, Dell) displayed graphical user interfaces (GUI).

The finger-surface contact forces and torques were measured via a 6D force sensor (Nano17 Titanium, ATI) placed on an acrylic plate fixed on the breadboard. We utilized the same sensor for all force measurements to ensure consistency in the gathered data. Above the sensor, we put another acrylic plate (3\,mm thickness) with a 3D-printed side to place the surfaces. The finger accelerations during active explorations were captured via a high-resolution, high-bandwidth, three-axis analog accelerometer (ADXL356, Analog Devices) attached to the index fingernail via double-sided tape. A data acquisition board (PCIe-6323, NI) collected the data from the force and acceleration sensors with a 10\,kHz sampling rate. The finger position was measured with a 2D infrared position sensor (NNAMC1580PCEV, Neonode) with 0.1~mm resolution and a 60~Hz sampling rate. 

For thermal data measurements, a miniature thermistor (223Fu3122-07U015, Semitec) was attached to the center of the participants' index finger pad with a 1~mm thin strip of insulation tape, and a heat flux sensor (FHF05-15X30, Hukseflux) was mounted on a 3D-printed frame. When touched with the index and middle fingers, the thermistor made contact with the material and measured the contact temperature, whereas the heat flux sensor lay under the middle finger. To ensure full coverage of the finger pad, we chose the FHF-05 15$\times$30~mm heat flux sensor. The initial surface temperatures were measured with an infrared thermometer. The data from the heat flux sensor and thermistor were sampled at 100~Hz using the data acquisition board (PCIe-6323, NI). 

A motorized linear stage (NRT100/M, Thorlabs) was equipped for the pressing measurements. A 3D-printed hand support was mounted on the linear stage to keep the participant's index finger in place and provide a 20~° contact angle relative to the surface. The finger holder had an adjustable bolt to accommodate different finger lengths. The pressing velocity was controlled by a stepper motor controller (BSC201, Thorlabs) along the vertical axis. The vertical configuration was achieved by using a right-angle bracket (NRT150P1/M, Thorlabs). The hand support and stage allowed the index finger to land on the center of the material sample.

The audio signals were recorded via a cardioid condenser microphone (AT2020 USB+, Audio Technica) in the 20~Hz-20~kHz range with a sampling frequency of 44.1~kHz. As cardioid microphones are unidirectional and most sensitive at the front, we placed the microphone front close to the interaction area.

The visual data was captured by two USB machine vision cameras (Alvium 1800 U-508c). These cameras recorded the interactions from both top and side views at 44~FPS with a resolution of 1200$\times$1200 pixels. The top-view camera was also used to take high-quality images of each material sample. To have enough material exploration space for participants, we placed the top-view and side-view cameras at 450~mm and 170~mm working distances. We chose the 16~mm C series fixed focal length lens (Techspec, Edmund Optics) to address these working distances. Additionally, we used an LED ring light (EFFI-RING, Effilux) to provide illumination in combination with a polarizer (EFFI-RING-POL, Effilux) to eliminate glare and suppress the reflections from the illumination. The two cameras and the LED ring light were mounted to custom-made, partly 3D-printed stands.

We implemented appropriate solutions for synchronization among the recorded multisensory data. A sound cue signaled the start of each experiment for participants. Subsequently, software triggers activated the data acquisition card and microphone, while hardware triggers initiated the cameras. Timestamps were assigned to samples from these devices, ensuring synchronous data recording.

\subsection{Data collection procedure} \label{sec:procedure}
\textbf{Finger-surface interaction data:} Two males, the first two authors of this paper, with an age of 26 years, participated in the finger-surface interaction data collection.

Both participants gave informed consent and agreed to share their finger-interaction data publicly. Before each measurement, they washed their hands with soap and dried them using a towel. Then, each participant underwent an instruction and training process. They sat on a chair and put their arm on the armrest, facing the monitor. Afterward, their finger-surface interactions were collected while they explored the surfaces by applying static contact, pressing, tapping, and sliding. The detailed procedures for each measurement are described in the following, and Supplementary Video 1 visualizes each interaction. Recording all types of physical interaction data from one participant for fifty surfaces took approximately (not consecutively) 24 hours.

\noindent\textit{a) Static contact:} In these measurements, our goal was quantifying thermal interactions occurring during finger-surface contact. Previous works on thermal perception and rendering determined two major thermal parameters for discriminating materials by touch as the heat flux conducted out of the skin and the corresponding skin temperature~\cite{ho2006contribution}. 

Following a similar approach to \cite{choi2018data}, we first measured the initial temperature of a surface sample via an infrared thermometer. Afterward, the thermistor was attached to the participant's index finger pad, whereas the 3D-printed frame with the heat flux sensor was positioned on the sample surface and then stabilized with weights. The measurement started with a sound cue while the participants' fingers lay just over the samples, enabling the thermistor to measure the initial finger temperature. After hearing a sound cue, the participant was instructed to place their index and middle finger on the frame. They maintained a constant contact force of 3~N for 60 seconds for each surface with both fingers to keep the contact area constant. 

\noindent \textit{b) Pressing:} With these measurements, we aimed to determine how different materials respond to pressure from the index finger by measuring the force-indentation depth relationship when pressed with the index finger. We chose to record normal force and indentation depth, as recent studies have shown that finger contact area alone is not a distinguishable metric when determining the compliance of objects, and kinesthetic cues likely augment our judgments of compliance~\cite{srinivasan1995tactual}. 

First, the participant mounted their right hand onto the hand support. Then, the linear stage moved in the vertical direction with the participant's finger until reaching 3~N normal force. After staying there for two seconds, the finger was moved to the start position.

\noindent \textit{c) Tapping:} By tapping measurements, we sought to capture the surface hardness. Previous works showed that we can discriminate hardness by tapping on a surface~\cite{culbertson2016importance}, and tap spectral centroid of the contact vibrations is a large contributor in material discrimination~\cite{richardson2022learning,vardar2019fingertip}. 

Before tapping measurements, the accelerometer was placed on the participant's fingernail. After the calibration process of the accelerometer, the participant tapped six times on the center of the surface sample using their dominant hand's index finger. The start of each experiment was indicated with a sound cue. They were instructed to apply a maximum force of 1~N in the first three taps and then use a force level that surpasses 2~N for the rest. The desired force level and real-time measured contact force were shown to participants via a Matlab GUI and the force sensor.

\noindent \textit{d) Sliding:}
In these measurements, we aimed to capture the friction and roughness properties of the surfaces. Since measured surface friction and roughness features change with applied force and exploration speed~\cite{lederman1999perceiving,culbertson2016importance}, our experiments comprehensively explored surfaces across a spectrum of speeds and applied forces.

Similar to tapping, the accelerometer was placed on the fingernail of the participant before these experiments. After the calibration process, the participant was instructed to follow a custom GUI shown on the monitor; refer to \href{https://www.sens3.net}{SENS3} website for visualization of GUI. The GUI showed a 5$\times$6 matrix of UI elements, each corresponding to a specific finger force and speed range, shown in red color initially. The force ranges of elements corresponded 0-0.2, 0.2-0.4, 0.4-0.6, 0.6-0.8, 0.8-1~N, whereas the speed ranges represented 0-33, 33-66, 66-99, 99-132, 132-165, and 165-200~mm/s. The maximum force and speed were limited to 1~N and 200~mm/s, respectively, as beyond values were challenging to maintain for long. After hearing a sound cue, the participant explored the surface through unconstrained sliding, as this method enables capturing the interaction data more efficiently than constrained exploration~\cite{culbertson2013generating}. When the participant achieved a specific force-speed combination, the color of the corresponding UI element shifted to green. Each force-speed range required an unconstrained finger-surface exploration with a duration of five seconds. The participants were free to achieve any force-speed combination randomly. The recording from a surface sample was completed when all the UI elements of the matrix turned green. The participant's applied force and speed were shown at the top of the GUI. \\

\noindent \textbf{Perceptual data:} Through the perceptual measurements, we aimed to record haptic psychophysical sensations felt by participants during the free multisensory exploration of surfaces using their dominant index finger. For this, we used the semantic differential method~\cite{Okamoto2013PsychophysicalDO} with 15 points, where participants rated the surfaces based on eight opposing adjectives by moving sliders (implemented in MATLAB's GUI) for each adjective. We selected a subset of surface adjectives used by~\cite{baumgartner2013visual}: rough--smooth, flat--bumpy, sticky--slippery, hot--cold, regular--irregular, fine--coarse, hard--soft, and wet--dry. Before the experiment, we asked the participants to wash and dry their fingers first and then explore all the surfaces without any sensory restrictions---they were free to interact with surfaces by seeing, hearing, and touching---to adjust their adjective limits mentally. Following this, we conducted mock trials with five random surfaces to acclimate the participants to the study. During the experiments, we randomly presented the surfaces to the participants. We asked participants to explore the surface freely using all their three senses (visual, audio, and haptic) for fifteen seconds after they heard the sound cue. Although they could interact with the surfaces as they wished, we encouraged them to use all haptic exploratory procedures explained during mock trials. After the exploration, the participants rated the surface feels by adjusting the scales in the GUI. The experimental procedure followed the Declaration of Helsinki and was approved by TU Delft's ethics committee with application number 3469. Three women and ten men with an average age of 26.84 years (standard deviation, SD: 2.034) participated in the experiment. All participants gave informed consent. The perceptual data collection procedure took 2 hours per participant.

\subsection{Collected data}
\href{https://www.sens3.net}{\emph{SENS3.net}} hosts our database. 
The website was designed to provide a comprehensive resource for the recorded data and the recording procedures. It consists of three pages:

\begin{enumerate}
    \item `Home' page: provides a general overview of the database, featuring informative figures and videos that enable users to grasp the essence of our database and the recording techniques.
    \item `Recording Setup' page: includes an overview of the recording setup, data recording procedures, GUI, hardware list, and a figure depicting the camera and microphone arrangement.
    \item `Surfaces' page: hosts the collection of finger-surface interaction data organized into ten distinct categories. Users can click on a category to view the photos of each surface within it and download the associated data. The surface data are further aggregated based on different participants. The `Surfaces' page also includes a `Metadata' section, offering details about available files along with their descriptions and further information about surface physical properties; see Table~\ref{tab:metadata}).
    \item `About' page: contains citation detail for the paper and additional information about the authors.
    
\end{enumerate}

\begin{table}[]
\centering
\caption{Metadata for the recorded data files. (num) represents the allocated number tag of the surface.}
\begin{tabular}{ll}
\hline
File                         & Description                                                                                                                                                                           \\ \hline
\multicolumn{2}{c}{\textbf{Tapping}}                                                                                                                                                                                 \\
forces.csv                   & \begin{tabular}[c]{@{}l@{}}contact forces and torques in 6 columns \\ - Fx, Fy, Fz, Tx, Ty, Tz {[}N and Nmm{]}\end{tabular}                                                           \\
accelerations.csv            & \begin{tabular}[c]{@{}l@{}}accelerations in 3 columns \\ -Ax, Ay and Az {[}g{]}\end{tabular}                                                                                          \\
audio.wav                    & audio data                                                                                                                                                                            \\
video(num)\_1.mp4            & side view video                                                                                                                                                                       \\
video(num)\_2.mp4            & top view video                                                                                                                                                                        \\ \hline
\multicolumn{2}{c}{\textbf{Pressing}}                                                                                                                                                                                \\
forces.csv                   & \begin{tabular}[c]{@{}l@{}}contact forces and torques in 6 columns \\ - Fx, Fy, Fz, Tx, Ty, Tz {[}N and Nmm{]}\end{tabular}                                                           \\
stageposition.csv            & position of the linear stage {[}mm{]}                                                                                                                                                 \\
video(num)\_1\_2mms.mp4      & side view video                                                                                                                                                                       \\
video(num)\_2\_2mms.mp4      & top view video                                                                                                                                                                        \\ \hline
\multicolumn{2}{c}{\textbf{Static Contact}}                                                                                                                                                                          \\
forces.csv                   & \begin{tabular}[c]{@{}l@{}}contact forces and torques in 6 columns \\ - Fx, Fy, Fz, Tx, Ty, Tz {[}N and Nmm{]}\end{tabular}                                                           \\
temperature.csv              & index finger temperature {[}°C{]}                                                                                                                                                     \\
material\_temperature.csv    & surface temperature of the materials {[}°C{]}                                                                                                                                         \\
heatflux.csv                 & \begin{tabular}[c]{@{}l@{}}heat flux between middle \\ finger and material {[}W/m$^2${]}\end{tabular}                                                                                 \\ \hline
\multicolumn{2}{c}{\textbf{Sliding}}                                                                                                                                                                                 \\
Material\_sensor\_(num).csv  & \begin{tabular}[c]{@{}l@{}}contact forces and torques in 6 columns \\ - Fx, Fy, Fz, Tx, Ty, Tz  {[}N and Nmm{]}, \\ accelerations in 3 columns, \\ Ax, Ay and Az {[}g{]}\end{tabular} \\
Material\_IR\_pos\_(num).csv & \begin{tabular}[c]{@{}l@{}}4 columns of data - elapsed time {[}s{]}, \\ speed {[}mm/s{]}, position X {[}mm{]}, \\ position Y {[}mm{]}\end{tabular}                                    \\
Material\_(num).mp4          & top view video                                                                                                                                                                        \\
Material\_(num).wav          & audio data                                                                                                                                                                            \\ \hline

\multicolumn{2}{c}{\textbf{Sensation rating
}} \\

Order\_par(num).csv & order of material presented to participant\\
Ratings\_par(num).csv & adjective ratings\\
Video(num).mp4 & isometric view video
\\ \hline
material(num).tif            & top view image                                                                                                                                                                        \\
thickness.csv                & thickness of the materials                                                                                                                                                            \\ \hline

\end{tabular}
\label{tab:metadata}
\end{table}

\begin{figure}[]
    \centering
    \includegraphics[width=\linewidth]{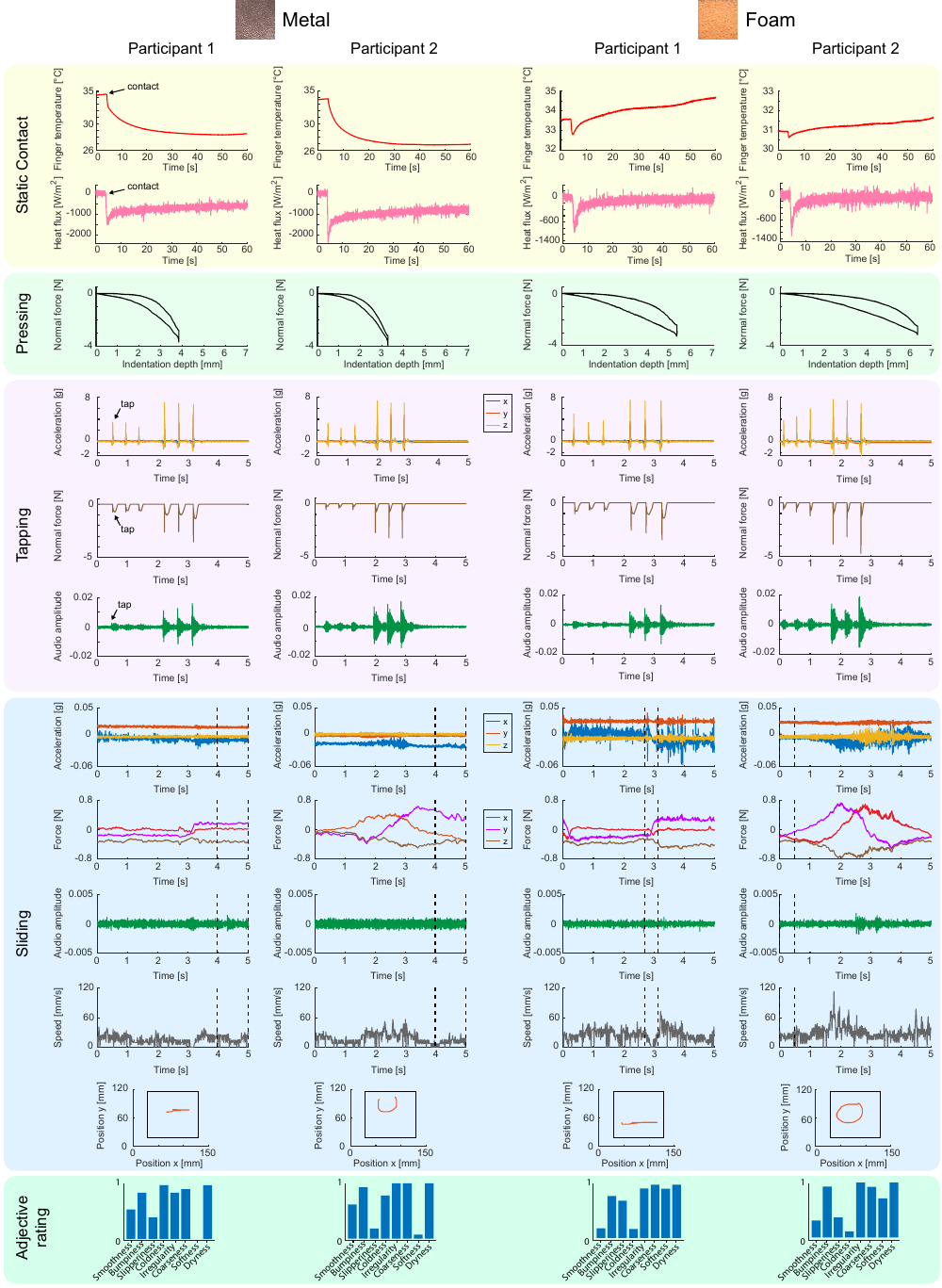}
    \vspace*{-8mm}
    \caption{Example recordings from two participants while interacting with two different surfaces: metal and foam. The sliding data comprises only a 5-second segment from the complete recording for a clear visualization. The dashed lines on the sliding plots highlight one [0.2-0.4 N, 0-33 mm/s] force-speed pair region for each of the measurements. The adjective ratings are normalized based on the maximum and minimum value reported by the participant across all surfaces.}
    \label{fig:data_plots}
\end{figure}

Fig.~\ref{fig:data_plots} showcases the physical data collected from two participants while interacting with two distinct surfaces (metal and foam) with static contact, pressing, tapping, and sliding (only a snippet of five seconds), as well as their adjective ratings collected from psychophysical experiments. The variations in data are visible for different surfaces and participants and demonstrate the significance of variety in surface and participants for the texture databases. For example, there are visible differences between changes in skin temperature, heat flux, compliance hysteresis, and contact accelerations between metal and foam. Similarly, the interaction data from both participants discriminate from each other; these differences could be caused by variations in their exploratory behavior (compare finger position graphs at the last row) and their skin properties~\cite{callier2015kinematics}. 

\section{Results of the perceptual experiments}\label{sec:results}

We used principal component analysis (PCA) to analyze the adjective ratings, first normalizing them using z-scores within each participant. The resulting normalized ratings were averaged across all participants, yielding a 50$\times$8 matrix (surfaces by adjective pairs). Subsequently, we computed the eigenvectors of the covariance of the normalized adjective matrix, sorting the diagonal elements in descending order to determine the optimal number of dimensions that capture the data's variance. Four dimensions were chosen as they represent 95\% of the total variance. Following this, we did factor analysis through varimax rotation of the component matrix~\cite{drewing2018systematic} to improve interpretability; see Table.~\ref{tab:PCA} for the rotated component matrix with four components and their corresponding factor loadings.

\begin{table}
\caption{Rotated Component Matrix}
\centering
\begin{tabular}{ccccc}
\hline
                             & \multicolumn{4}{c}{Principal Components}                                                                                            \\ \cline{2-5} 
\multirow{-2}{*}{Adjectives} & 1                               & 2                              & 3                              & 4                               \\ \hline
Rough--Smooth                 & \cellcolor[HTML]{C0C0C0}-0.7878 & -0.0689                        & 0.1380                         & 0.5593                          \\
Flat--Bumpy                   & \cellcolor[HTML]{C0C0C0}0.9263  & 0.2261                         & 0.1576                         & -0.1397                         \\
Sticky--Slippery              & 0.0183                          & -0.0629                        & \cellcolor[HTML]{C0C0C0}0.9605 & -0.1643                         \\
Warm--Cold                    & -0.2606                         & -0.4775                        & -0.1735                        & \cellcolor[HTML]{C0C0C0}0.7923  \\
Regular--Irregular            & \cellcolor[HTML]{C0C0C0}0.9246  & 0.1174                         & -0.0213                        & -0.1153                         \\
Fine--Coarse                  & \cellcolor[HTML]{C0C0C0}0.9237  & 0.1741                         & 0.0140                         & -0.3093                         \\
Hard--Soft                    & 0.2604                          & \cellcolor[HTML]{C0C0C0}0.8683 & -0.1229                        & -0.3898                         \\
Wet--Dry                      & 0.2543                          & 0.2105                         & 0.1923                         & \cellcolor[HTML]{C0C0C0}-0.8912 \\ \hline
\end{tabular}
\label{tab:PCA}
\end{table}

Table.~\ref{tab:PCA} shows that the first rotated component, accounting for 66\% of the variance, encompasses adjective ratings such as "rough--smooth," "flat--bumpy," "regular--irregular," and "fine--coarse," indicating roughness cue. Similarly, the second rotated component, explaining 16\% of the total variance, is characterized by the "hard--soft" adjective rating, signaling compliance cue. The third rotated component, describing 9\% of the total variance, is represented by the "sticky--slippery" adjective rating, indicating the friction cue. Lastly, the fourth rotated component, capturing 4\% of the total variance, is associated with adjective ratings such as "warm--cold" and "wet--dry," hinting at thermal cues.
\newpage
\section{Discussion}\label{sec:discussion}
We presented \href{https://www.sens3.net}{SENS3}, an open-source multisensory database of finger-surface interactions and corresponding psychophysical sensations. This database contains recordings of when two participants interacted with 50 surfaces by static contact, pressing, tapping, and sliding. Our database captures a wide range of sensory information, such as surface photographs, video and sounds of the interactions, contact forces and torques, finger accelerations and positions, skin temperature, and heat flux transfer. In addition to the physical recordings, we captured the psychophysical sensations generated during free exploration through adjective ratings given by thirteen participants. We hope that \href{https://www.sens3.net}{\emph{SENS3}} will provide the necessary data for content design for multisensory user interfaces, understanding human touch and multisensory integration, and providing robots with humanlike touch sensations. 

Despite our diligent efforts, the current dataset version exhibits some limitations. It comprises finger-surface interaction recordings from two trained male participants and perceptual ratings from thirteen participants, all collected by exploring fifty surfaces. This limited participant pool hampers our ability to capture the full spectrum of finger biomechanics and human perception. Furthermore, the surface diversity within the dataset represents merely a subset of surfaces encountered in daily life. Additionally, separating physical interaction recordings and adjective ratings poses a challenge in establishing direct correlations between sensations and physical data. Given the interconnected nature of adjective ratings and the inability of a single exploratory method to encapsulate all these sensations, we opted for free exploration, integrating all four exploratory procedures, to gather comprehensive perceptual data~\cite{okamoto2012psychophysical,cavdan2021task} and high-quality finger-surface interaction data separately with distinct exploratory procedure. 

We designed SENS3 as open-ended so that we can address the shortcomings mentioned above. We envision augmenting the dataset with data collected from a more diverse range of surfaces and participants, encompassing variations in age, gender, and finger biomechanics. Our expansion plans also incorporate surface characteristics, finger biomechanical measurements, and simultaneous perceptual data collection separately during each exploratory procedure. 

We aim to make \href{https://www.sens3.net}{\emph{SENS3.net}} an open-source platform intended to be a collaborative hub where laboratories, companies, or designers from diverse backgrounds add their new collected data or texture rendering models designed for various interfaces, such as touch-based devices and wearables. This collective effort will accelerate progress in the field of multisensory user interfaces but also empower a wide range of applications, from enhancing accessibility for individuals with sensory impairments to revolutionizing entertainment experiences.

\section*{Acknowledgements}
This publication is part of the project ``From signal-based modeling to sensation-based modeling'' with project number 19153 of the research programme Veni partly financed by the Dutch Research Council (NWO) and Huawei Technologies. JK. Balasubramanian and B. Kodak contributed equally to this work. 
\bibliographystyle{splncs04}
\bibliography{bibli.bib}
\end{document}